\documentclass[prd,reprint,showpacs,showkeys]{revtex4}
\usepackage{graphics}
\usepackage{eurosym}
\usepackage{amsfonts}
\usepackage{amssymb}
\usepackage{amsmath}
\usepackage{graphicx}
\usepackage[font={footnotesize,it}]{caption}
\usepackage[colorlinks=false,
            linkcolor=red,
            urlcolor=red,
            citecolor=blue]{hyperref}

\usepackage{color}

\begin{document}

\title{Charged  Vaidya-Tikekar model for super compact star}

\author{Jitendra Kumar \& Amit Kumar Prasad}
\email{jitendark@gmail.com}
\affiliation{Department of Applied Mathematics, Central University of Jharkhand, Ranchi-835205, India.}

\author{S. K. Maurya}
\email{sunil@unizwa.edu.om}
\affiliation{Department of Mathematical and Physical Sciences,
College of Arts and Science, University of Nizwa, Nizwa, Sultanate of Oman}

\author{ Ayan Banerjee}
\email{ayan\_7575@yahoo.co.in}
\affiliation{Astrophysics and Cosmology Research Unit, University of KwaZulu Natal, Private Bag X54001, Durban 4000,
South Africa.}

\date{\today }

\begin{abstract}
 In this work, we explore a class of compact charged spheres that have been tested against experimental and observational constraints with some known compact stars candidates.  The study is performed by considering the self-gravitating, charged, isotropic fluids which is more pliability in solving the Einstein-Maxwell equations.  In order to determine the interior geometry, we  utilize the Vaidya-Tikekar \cite{Vaidya} geometry for the metric potential with Riessner-Nordstrom metric as an exterior solution. In this models, we determine constants after selecting some particular values of M and R, for the compact objects SAX J1808.4-3658, Her X-1 and   4U 1538-52. The most striking consequence is that hydrostatic equilibrium is maintained for different forces, and the situation is clarified by using the generalized Tolman-Oppenheimer-Volkoff (TOV) equation. In addition to this, we also present the energy conditions, speeds of sound and compactness of stars that are very much compatible to that for a physically acceptable stellar model. Arising solutions are also compared with graphical representations that provide strong evidences for more realistic and viable models,  both at theoretical and astrophysical scale.
\end{abstract}

\keywords{Einstein-Maxwell System; Perfect Fluids;  Compact Stars}

\maketitle

\section{Introduction}

After the discovery of general relativity in 1915 by Einstein, it's became an important tool for understanding and explaining the gravitational system. In particular, obtaining a singularity free interior solution for compact astrophysical objects is an important issue in relativistic astrophysics for the past two and a half decades. At some points, we usually refer compact objects as a collectively of different types of high density objects like white dwarfs, neutron stars and quark stars, that form at the end of their stellar evolution. Therefore, in order to study the structure of such stars form microscopic composition and properties of dense matter in an extreme condition is one of the most fundamental problems in modern astrophysics.  However, in spite of the fact that such extreme densities nuclear matter may consist not only of leptons and nucleons but also consists of some mesons, hyperons, baryon resonances as well as strange quark matter (SQM), in their different forms and phases at the time of stellar evaluation. Indeed, works that reviewed by different authors demonstrate the difficulties associated with constructing solutions for obtaining a comprehensive description of compact objects.

From the above description one can easily understand that obtaining a reliable description of dense matter compact object is not an easy task, though there are several theoretical investigations, laboratory experiments as well as observations test have been performed during the last few decades. But the observational data form compact stars may soon provide some  information about largest uncertainties in nuclear physics that rely heavily on the equations of state (EoS) at nuclear and supranuclear densities. In general, this could be achieved by estimating their mass and radius \cite{Mak,Burikham,Boehmer1} which depends on EoS \cite{Ray,Negreiros,Varela,Maharaj1}. The motivation to undertake such a task was initiated by the discovery of the pulsar PSR J1614-2230 that produced by hot spots on the surface of compact stars \cite{Hessels}  led to several interpretative problems such as neutron stars (NSs) and quark stars (QSs). However, this pulsar profile does not depend only on  the stellar mass and radius,  but also depends on several other features like moment of inertia, quadrupole moment and higher multi-pole moments etc. According to recent observation there are many compact objects, namely,  X-ray burster 4U 1820-30, X-ray pulsar Her X-1, X-ray sources 4U 1728-34, PSR 0943+10 and RX J185635-3754, whose masses and radii are not compatible with the standard neutron star models. The problem is that we still have lack of information about the nuclear matter density, so theoretical studies hints that pressure is likely to be anisotropic within the stellar radii, i.e.,  the radial pressure and the tangential pressure. The search for anisotropic superdense stars were initiated by Bowers and Liang \cite{Bower} and then a number of articles appeared related with this in \cite{Kalam1,Kalam2,Rahaman1,Bhar1,Bhar2,Banerjee1,Banerjee2,Ratanpal,Thirukkanesh}.

The search for an exact solutions of Einstein field equations for static isotropic and anisotropic astrophysical objects are excellent testbeds with growing interest to mathematician as well as physicists. However, most of the exact interior solutions for both isotropic and anisotropic cases do not  satisfy the general physical required conditions of the stellar systems. Therefore, exact solutions of Einstein-Maxwell field equations are also important in relativistic astrophysics. The conjecture is the obtained solution may be utilized to model for a charged relativistic star which match to the Reissner-Nordstr$\ddot{\text{o}}$m  exterior spacetime at the boundary. In this context one can avoid the gravitational collapse of a spherically symmetric distribution of matter, because of gravitational attraction is counterbalanced by the repulsive Coulombian force in addition to the pressure gradient. In this connection, Ivanov \cite{Ivanov} and Sharma \textit{et al.} \cite{S1} showed that due presence of the electric field affects the nature of  luminosities, redshifts and maximum masses of  relativistic star. Later on, Takisa and Maharaj \cite{Takisa} have proposed an exact solutions for charged anisotropic polytropic spheres. Imposing different equations of state some of these charged solutions have studied in \cite{Jose1,Jose2,Maurya3,Maharaj2,Barreto1,Barreto2}.

The simplest known procedure that can be added for constructing a static charged perfect fluid interior solution for compact objects are either to prescribe a metric ansatz or an equation of state relating pressure and density. In the present article we consider a well known ansatz for one of the metric functions, namely, Vaidya and Tikekar \cite{Vaidya}, who prescribed an ansatz for the geometry of the t = \textit{constant} hypersurface. The main important feature is that the solution characterizes a class of static spherically symmetric perfect fluid configuration and provides an exact solution of Einstein’s equations.  Following this technique, a large number of solutions have been studied in \cite{10,11,12,13,Patel}. In a recent treatment Naveen and Bijalwan \cite{Bijalwan,Bijalwan1} have obtained a charged perfect fluid model with generalized electric intensity
for all $K$  except for $0< K < 1$ and extending this point of view
Kumar and Gupta \cite{2013,2014} obtained another solution for $0<K<1$. 
In the present problem we focused on a charged
fluid sphere starting with Vaidya and Tikekar \cite{Vaidya} metric potential and 
tested our model with some standard observed mass
and radius of the compact stars candidates as proposed in \cite{Gangopadhyay}.

The paper is organized as follows:  following a brief introduction in Sec. \textbf{I}, then we give the general relativistic formulation of Einstein-Maxwell system of equations for a relativistic stellar model in Sec. \textbf{II}. Paying particular attention to solve the system of equations analytically, we assume a particular form of  metric potential, namely, Vaidya-Tikekar to generate exact solutions and obtain the expression for energy density and pressure in the same section. Then, we match the interior charged fluid to the exterior Reissner-Nordstr$\ddot{\text{o}}$m line element in Sec. \textbf{III}. Next, in Sec. \textbf{IV}, we have discussed briefly some physical features of the proposed model maintaining  the regularity conditions and obtained results are compared with observational data. Finally in Sec. \textbf{V}, we give a brief discussion.

\section{Einstein field equations}

In this work, we consider the  static spherically symmetric spacetime for seeking solution
of a compact stellar object, that can be written in Schwarzschild coordinates as \begin{eqnarray}
 ds^{2} = e^{\nu(r)}dt^{2}-e^{\lambda(r)}dr^{2} -r^{2}(d\theta^{2}+\sin^{2}\theta d\phi^{2}),\label{1}
\end{eqnarray}
where the unknowns $\nu(r)$ and $\lambda(r)$ are both metric functions in
terms of radial coordinate, which yet to be determined by solving the field equations.

The most important question that arise is the matter distribution inside a compact star. Here, we consider the case of a charged gravitating object with isotropic pressures. Then the energy momentum tensor $T^{i}_{j}$
will include the terms from the Maxwell's equation $E^{i}_{j}$ and the complete form  of energy-momentum tensor is
\begin{eqnarray}
T^{i}_{j}+E^{i}_{j}= \big[(c^{2}\rho+p)u^{i}u_{j}-p\delta^{i}_{j}+\dfrac{1}{4\pi}(-F^{im}F_{jm}
+\dfrac{1}{4}\delta^{i}_{j}F_{mn}F^{mn})\big], \label{2}
\end{eqnarray}
where $u_j$, and ${\delta^i_j}$ stand for 4-velocity of the fluid,
 and the metric tensor, respectively. Generally first component standards
 for energy momentum tensor; e.g. for a perfect  fluid we have
 $\left[(c^{2}\rho+p)u^{i}u_{j}-p\delta^{i}_{j}\right]$, where $\rho$ is the matter
 density and $p$ is the pressure of the fluid. The basic argument to assume the perfect  fluid implies that the  flow of matter is adiabatic, no heat flow, radiation, or viscosity is present \cite{Misner}. The second term associate with electromagnetic stress-energy tensor from the Maxwell's field equation and hence they will follow the relation
$\left[\sqrt{-g}F^{i j}\right]_{,j}=4\pi j^{i}\sqrt{-g}$, where
$F_{ij}$ denote the skew symmetric electromagnetic field tensor and $j^{i}$ is the four-current density.

 Now, accomplishing the effects due to the electric field and pressure isotropy, and using the
metric (\ref{1}) with stress tensor given in Eq. (\ref{2}), the Einstein field equation, $G_{\mu\nu}$ = -$\kappa \left[T^{\mu}_{\nu}+E^{\mu}_{\nu}\right]$,
where $R^{i}_{j}-\dfrac{1}{2}R\delta^{i}_{j}$, provides the following relationships
 \begin{eqnarray}
 \dfrac{\lambda'}{r}e^{-\lambda}+\dfrac{(1-e^{-\lambda})}{r^{2}}=\kappa \rho+\dfrac{q^{2}}{r^{4}},\label{3}\\
\dfrac{\nu'}{r}e^{-\lambda}-\dfrac{(1-e^{-\lambda})}{r^{2}}=\kappa p-\dfrac{q^{2}}{r^{4}},\label{4} \\
\bigg(\dfrac{\nu''}{2}-\dfrac{\lambda'\nu'}{4}+\dfrac{\nu'^{2}}{4}+\dfrac{\nu'-\lambda'}{2r}\bigg)e^{-\lambda}=\kappa p+\dfrac{q^{2}}{r^{4}},\label{5}
\end{eqnarray}
with $ \kappa= 8\pi$ (geometrized units G = c = 1) and prime denotes the differentiation with
respect to the radial coordinate. The total charge inside a radius r is given by
\begin{eqnarray}
 q(r)= r^{2}\sqrt{-F_{14}F^{14}}=r^{2}F^{41}e^{(\lambda+\nu)/2} = 4\pi\int_{0}^{r}\sigma r^{2}e^{\lambda/2}dr.\label{6}
 \end{eqnarray}
Also $ F_{14}$ is the only non-vanishing component of the skew-symmetric electromagnetic tensor i.e.,
$ F_{14}= -F_{41}$. Equations (\ref{3})-(\ref{5}) are invariant under the transformation q(r) = - q(r) and
$\sigma = -\sigma$. Here, we exclusively deal with the positive square root of $q^2$.

 \par Note that Eqs. (\ref{3})-(\ref{5}) provide three independent equations,
 for five unknown quantities i.e. $ \nu, \lambda, \rho( r), p(r)$ and $q(r)$ which we have to solve
simultaneously to get our results. As, obtaining an explicit solutions to the
Einstein field equations is a difficult tusk due to highly nonlinearity of the
equations. Thus, we will reduce the number of unknown functions by assuming a well
known form of metric potential \cite{Vaidya}
 \begin{eqnarray}
 e^{\lambda}=\dfrac{K(1+Cr^{2})}{K+Cr^{2}},~~~~0<K<1, \label{7}
 \end{eqnarray}
 where $C =-K/R^2$, and $K$ \& R are two parameters which characterize the geometry of the star. Our choice of  Vaidya-Tikekar ansatz is physically well motivated and has been studied for uncharged superdense stars by Tikekar \cite{Tikekar} and Maharaj and Leach \cite{Leach}. This facilitated the model in an interesting geometric meaning as
deviation from sphericity of 3-space geometry. It may also be noted that metric
potential restricts the geometry of the 3-dimensional hypersurfaces {t = \emph{const.}}
to be spheroidal and when K = 0 the hypersurfaces {t = \emph{const}.} become spherical.

The solution of EFEs (\ref{3})-(\ref{5}), is in a different but equivalent form if we introduce
$Cr^{2}=x $ and $ e^{\nu}=Z^{2} $ as a new variables. After a little bit algebraic calculation we have
 \begin{eqnarray}
 \dfrac{(K-1)(3+x)}{K(1+x)^{2}}-\dfrac{q^{2}C}{x^{2}}=\dfrac{\kappa c^{2}\rho}{C},\label{8}\\
Z''-Z'\Bigg(\dfrac{\sqrt{C}K}{\sqrt{x}(K+x)}+\dfrac{\sqrt{xC}}{(1+x)}\Bigg)+\dfrac{C}{(\sqrt{K+x})}\times \nonumber \\\Bigg(\dfrac{(K-1)x}{(1+x)}-2q^{2}CK(1+x) \Bigg)Z= 0, \label{9} \\
\dfrac{(K+x)}{\sqrt{xC}K(1+x)}\dfrac{2Z'}{Z}+\dfrac{(1-K)}{K(1+x)}+\dfrac{q^{2}C}{x^{2}}= \dfrac{\kappa p}{C},\label{10}
  \end{eqnarray}
where  prime denote differentiation with respect to the variable x. 
Furthermore, to transform the field equations to a more convenient
form we introduce another variables defined by
\begin{eqnarray}
 Y=\sqrt{\dfrac{K+x}{1-K}},~~~~~\text{and}~~~~~  Z=(1+Y^{2})^{1/4}\Phi, \label{11}
 \end{eqnarray}
and plugging the values of $Y $ and $ Z $ into the equation (\ref{9}), we get 
\begin{eqnarray}
\dfrac{d^{2}\Phi}{dY^{2}}+\chi\Phi=0\label{12}
 \end{eqnarray}
 where for notational simplicity we use
 \begin{eqnarray} 
 \chi=-\dfrac{1}{(1+Y^{2})}\Bigg(K-1-2K q^{2}\dfrac{(1+x^{2})C}{x^{3}}+\dfrac{2-3Y^{2}}{4(1+Y^{2})^{2}}\Bigg).\label{13}
 \end{eqnarray}
In order to solve the second order differential equation
 (\ref{12}) more easily, we have chosen  $ \chi=-2\theta^{2}/(Y^{2}\theta^{2}+b)$,
where $ \theta $ is a positive constant, and comparing with
Eq. (\ref{13}) leads to defining the total charge of the system as
 \begin{eqnarray}
 \dfrac{q^{2}}{x^{2}}=\dfrac{x}{2CK(1+x)^{2}}\bigg[K-1+\dfrac{2-3Y^{2}}{4(1+Y^{2})^{2}}+
 \dfrac{2\theta^{2}(1+Y^{2})}{(Y^{2}\theta^{2}+b)}\bigg]. \label{14}
 \end{eqnarray}
 
Moreover, using the  value of $ \chi $ from (\ref{13}) into the Eq. (\ref{12}), which yields
\begin{eqnarray} 
(Y^{2}\theta^{2}+b)\dfrac{d^{2}\Phi}{dY^{2}}-2\theta^{2}\Phi=0 ,\label{15} 
\end{eqnarray}
 Now, comparing Eq. (\ref{15}) with a known standard differential equation
 \begin{eqnarray}
 P_{0}\dfrac{d^{2}Y}{dX^{2}}+P_{1}\dfrac{dY}{dX}+P_{2}X = R,  \label{16} 
 \end{eqnarray} 
which leads to the following relations
$ P_{0} = Y^{2}\theta^{2}+b,~~ P_{1}=0,~~ P_{2} = -2\theta^{2},~~R=0$. Now, it is clear that the differential equation is exact and rearranging the terms one can write this in a conventional form $P_{2}-\dfrac{dP_{2}}{dY}+\dfrac{d^{2}P_{0}}{dY^{2}}=0$. Hence the primitive of the given equation is 
 \begin{center}
 $ P_{0}\dfrac{d\Phi}{dY}+(P_{1}-P'_{0})\Phi=\int R dY +A$,\\
 \begin{eqnarray}
 \implies \dfrac{d\Phi}{dY}-\dfrac{2Y\theta^{2}+1}{Y^{2}\theta^{2}+b}\Phi=\dfrac{A}{Y^{2}\theta^{2}+b},\label{17}
 \end{eqnarray}
 \end{center}
 where $ A $ and b stands for arbitrary constants. Finally,  the differential equation (\ref{17}) leads to the following expression
 \begin{eqnarray}\Phi(Y)=(Y^{2}\theta^{2}+b)\dfrac{A}{2\theta b^{3/2}}\bigg[\arctan\bigg(\dfrac{Y\theta}{\sqrt{b}}\bigg) +\dfrac{1}{2}\sin2\bigg(\arctan\bigg(\dfrac{Y\theta}{\sqrt{b}}\bigg)\bigg)\bigg]+B(Y^{2}\theta^{2}+b),  \label{18}
 \end{eqnarray}
 Now, using the relation (\ref{11}) for $Z$, which yields
 \begin{eqnarray}
 Z = A(1+Y^{2})^{1/4}\Bigg[(Y^{2}\theta^{2}+b)\dfrac{1}{\theta b^{3/2}}H(Y)+B(Y^{2}\theta^{2}+b)\Bigg],\label{19}
 \end{eqnarray}
 where $ H(Y)=\bigg[\dfrac{1}{2}\arctan\bigg(\dfrac{Y\theta}{\sqrt{b}}\bigg) +\dfrac{1}{4}\sin2\bigg(\arctan\bigg(\dfrac{Y\theta}{\sqrt{b}}\bigg)\bigg)\bigg]. $

 In addition, taking into account Eqs. (\ref{14}) and (\ref{19}) into equations (\ref{8})-(\ref{10}), it is straightforward to achieve the following nonzero components
of field equations
\begin{eqnarray}
\dfrac{\kappa c^{2}\rho}{C}=\dfrac{(K-1)(3+x)}{K(1+x)^{2}}-\dfrac{x}{2K(1+x)^{2}}\bigg[K-1+\dfrac{2-3Y^{2}}{4(1+Y^{2})^{2}}+\dfrac{2\theta^{2}(1+Y^{2})}{(Y^{2}\theta^{2}+b)}\bigg],\label{20}\\ \nonumber\\
\dfrac{\kappa p}{C}=\dfrac{Y^{2}x}{K(1+Y^{2})}\left[\dfrac{ \dfrac{N1\times N2}{(1-K)}+N3}{N4\times N1}\right]-\dfrac{1}{K(1+Y^{2})}+ N6,\label{21}
\end{eqnarray}   
where, $ N1=\dfrac{1}{\theta b^{3/2}}\left[\dfrac{1}{2}\arctan\bigg( \dfrac{Y}{\sqrt{b}}\theta\bigg)+\dfrac{1}{4}\sin2\bigg(\arctan\bigg( \dfrac{Y}{\sqrt{b}}\theta\bigg)\bigg)\right]+\dfrac{B}{A},\\ \\
N2=\dfrac{(Y^{2}\theta^{2}+b)}{2(1+Y^{2})^{3/4}}+ 2\theta^{2}(1+Y^{2})^{1/4},~~~~~~ N3=\dfrac{(1+Y^{2})^{1/4}}{2b(1-K)Y}\left[1+\cos2\bigg(\arctan\bigg( \dfrac{Y}{\sqrt{b}}\theta\bigg)\bigg)\right],\\ \\
N4=(1+Y^{2})^{1/4}(Y^{2}\theta^{2}+b),~~~N6=\dfrac{x}{2K(1+x)^{2}}\bigg[K-1+\dfrac{2-3Y^{2}}{4(1+Y^{2})^{2}}+\dfrac{2\theta^{2}(1+Y^{2})}{(Y^{2}\theta^{2}+b)}\bigg]. $\\ \\

\begin{figure}[h!]
\centering
\includegraphics[width=7cm]{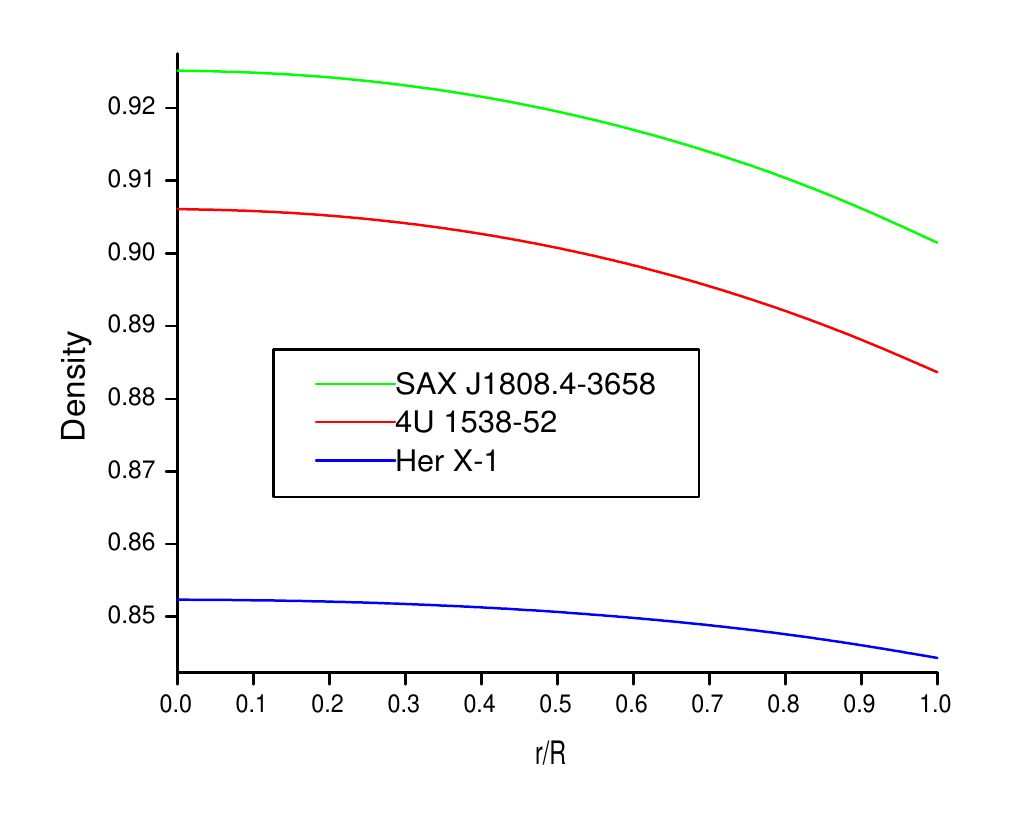} \includegraphics[width=7cm]{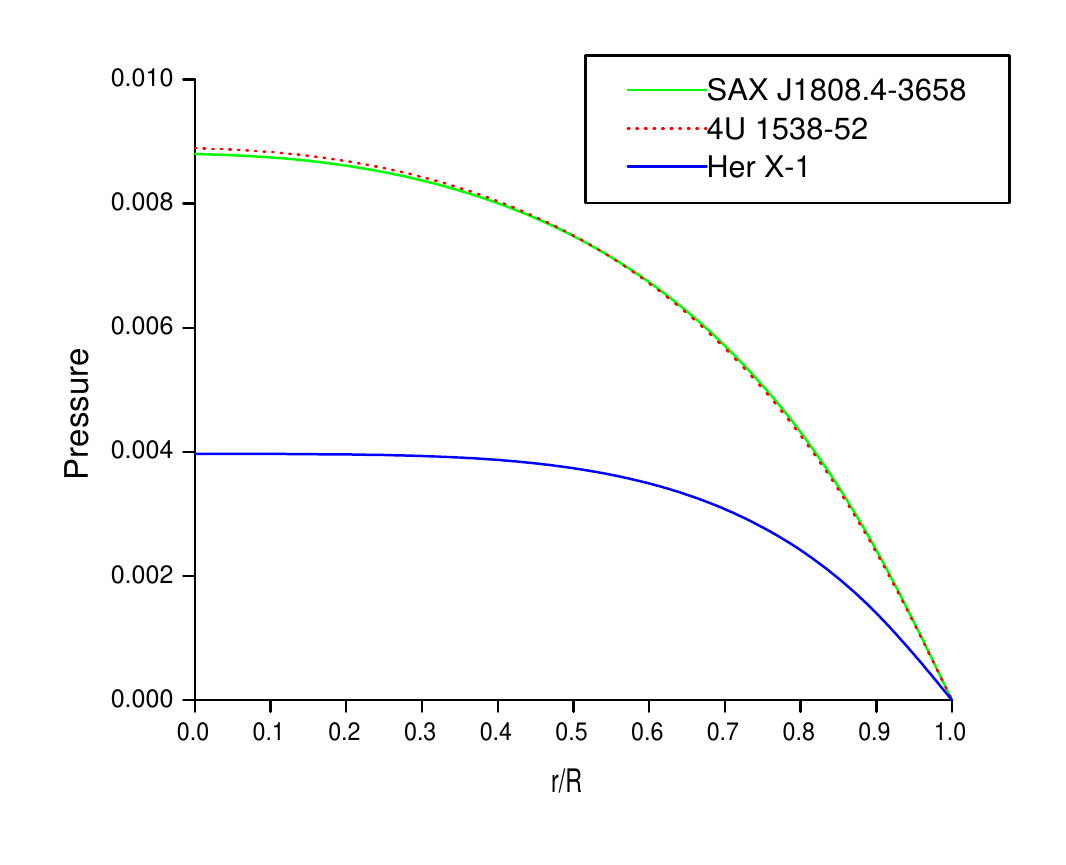}\\
\includegraphics[width=7cm]{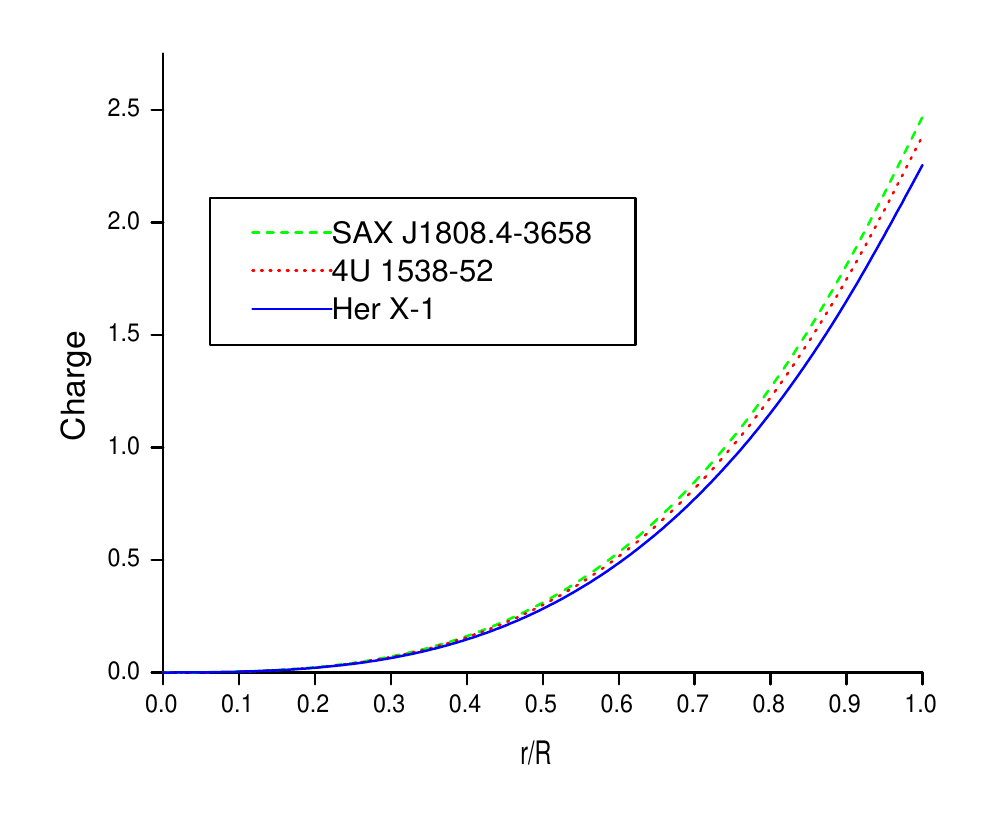} \includegraphics[width=7cm]{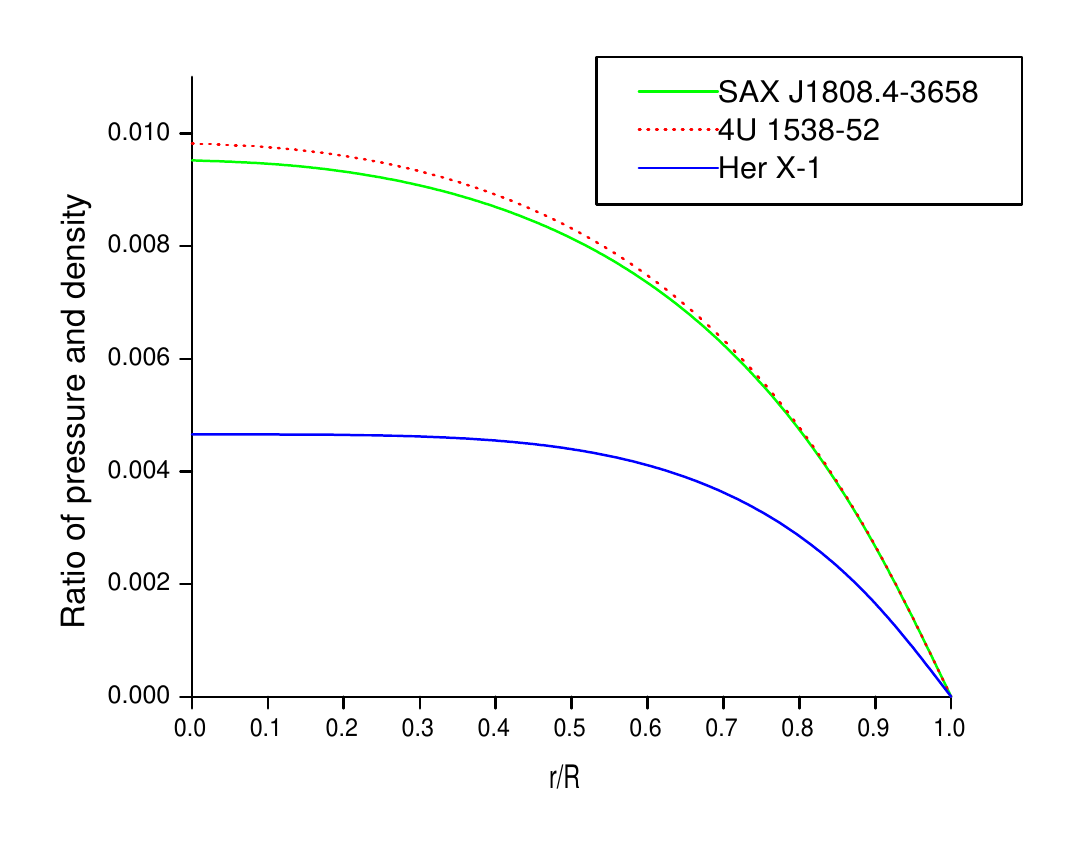}
\caption{\emph{The energy density,  pressure, charge and density-pressure in their normalized forms as a function of the radial coordinate are shown on the panels from top to bottom for the compact star candidates SAX J1808.4-3658, 4U 1538-52 and Her X-1.  For plotting,  we choose the values of physical parameters and constants are as follows: (i) K = 0.0032, b = 0.01, C = - 0.00099, $ \theta^{2} $ = 1.006 M = $0.9 M_{\odot}$, and R = 7.951 Km for SAX J1808.4-3658 (Table-I),
(ii) K = 0.0033, b = 0.05, C = -0.001, $\theta^{2}$ = 4.82, M = $0.87 M_{\odot}$ and R = 7.866 Km for 4U 1538-52 (Table-II), 
(iii) K = 0.034, b = 0.09, C=-0.01, $\theta^{2}$ = 0.835, M = $0.85 M_{\odot}$ and R = 8.1 Km for Her X-1 (Table-III).}}
\label{f1}
\end{figure}

With the purpose of determining the non-zero components, we are in a position
to determine the internal structure of compact stars. 
For compact star described by electrically charged fluid (within a certain radius) should comply with the following requirements throughout the interior radius:
\begin{itemize}
\item The energy density and pressure should be positive and regular within the radius.
\item $ (d\rho/dr)_{r=0}=0 $ and $ (d^{2}\rho/dr^{2})_{r=0}<0, $ so that gradient of density $ d\rho/dr $ is negative within $ 0<r < R $.
\item $ (dp/dr)_{r=0}=0 $ and $ (d^{2}p/dr^{2})_{r=0}< 0, $ so that pressure gradient $ dp/dr $ is negative within $ 0< r < R $.
\end{itemize}
The above three conditions imply that pressure and density should be maximum at the center and monotonically decreasing towards the surface, which is clear from Fig. (\ref{f1}). We are concerned here with charged isotropic case and according to Thirukkanesh and Mahara \cite{Marahaj1} charge distributions are singular at the origin, where the electric field does not vanish. That means, at the origin all our sources have vanishing electric field and finite proper charge density. From Fig. (\ref{f1}), it is clear that the charge will increase with the increasing charge fraction.

\section{Boundary conditions}
At this stage the interior solution is smoothly connected to the vacuum exterior Reissner-Nordstr$\ddot{\text{o}}$m metric at the junction surface with radius r = R. It is important to note that these three constants A, b and $\theta$  are fixed by suitable junction conditions imposed on the internal and external metrics at the hyper-surface. The exterior metric is given by 
 \begin{eqnarray}
 ds^{2}= \bigg( 1-\dfrac{2M}{r}+\dfrac{Q^{2}}{r^{2}}\bigg)dt^{2}
 -\bigg(1-\dfrac{2M}{r}+\dfrac{Q^{2}}{r^{2}}\bigg)^{-1}dr^{2}-r^{2}(d\theta^{2}+\sin^{2}\theta d\phi^{2}),\label{22}
\end{eqnarray}
where $ M $ is the total gravitational mass of the fluid distribution and is defined by
\begin{eqnarray} 
M=\zeta(R)+\xi(R), \label{23}
\end{eqnarray} 
with the definition $\zeta(R) = \dfrac{\kappa}{2}\int_{0}^{R}\rho r^{2}dr$, $\xi(R)= \dfrac{\kappa}{2} \int_{0}^{R}r\sigma qe^{\lambda/2}dr$ and Q = q(R)
represents, the mass within the sphere, the mass equivalence of the electromagnetic energy of distribution and Q is the total charge inside the sphere as suggested by
\cite{Florides}.
To do the matching properly, at the boundary surface r = R, we start by imposing the junction condition that the metric should be continuous \cite{Synge}. Thus, by joining the interior metric function $g_{rr}$ = $e^{\lambda}$  and $g_{tt}$ = $e^\nu$ with the
metric coefficient of the exterior Reissner-Nordstr$\ddot{\text{o}}$m spacetime given in (\ref{22}), we obtain the following conditions by using the continuity
 \begin{eqnarray} 
 e^{\lambda}=1-\dfrac{2M}{R}+\dfrac{Q^{2}}{R^{2}}, ~~~ \text{and}~~~ 
 y^{2}=1-\dfrac{2M}{R}+\dfrac{Q^{2}}{R^{2}},\label{24}
 \\p(R)=0,\label{27}~~~ \text{and}~~~ q(R)=Q.\label{25}
\end{eqnarray}
Now, using the conditions (\ref{24}) and (\ref{25}), we can fix the values of arbitrary constants. We being here, for some particular values of M and R, and the corresponding values of constant coefficients A, b and $\theta$ are determined. We use three configurations of stellar bodies, i.e., SAX J1808.4-3658, 4U 1538-52 and Her X-1 of masses 0.87 M($ M_{\odot} $), 0.85 M($ M_{\odot} $), and 0.9 M($ M_{\odot} $), respectively. Some possibilities of such types are tabulated in Table -4.

We will demonstrate here the interior and surface gravitational redshift $z_S$ of these compact sources by using the definition $z_S$ = $\Delta \lambda/\lambda_{e}$ = $\frac{\lambda_{0}-\lambda_{e}}{\lambda_{e}}$, where $\lambda_{e}$ is the emitted wavelength at the surface of a nonrotating star and $\lambda_{0}$ is the observed wavelength received at radial coordinate r.  Thus, one of these quantities is defined according to
\begin{eqnarray}
\label{eq26}
z_S = -1+\arrowvert g_{tt}(r) \arrowvert ^{-1/2} = -1+\left(1-\frac{2M}{R}+\dfrac{Q^{2}}{R^{2}}\right)^{-1/2},
\end{eqnarray}
where $g_{tt}(r)$ = $e^{\nu(R)}$ =$\left(1-\frac{2M}{R}+\dfrac{Q^{2}}{R^{2}}\right)$ is the metric function. Note that interior redshift  should decrease with the increase of radius and less than the universal bounds, found when different energy conditions holds. In the isotropic case gravitational redshift for  perfect fluid spheres is given by  $z_{s} <$ 2 \cite{Buchdahl,Straumann}. For an anisotropic star, this value admits higher redshifts,  $Z_{s}$ = 3.84, as given in Ref. \cite{Karmakar,Barraco}, but in the presence of cosmological constant provides a significant increase up to $z_{s} \leq$ 5 \cite{Boehmer}, which is consistent with the bound  $z_{s} \leq$ 5.211 obtained by Ivanov \cite{Ivanov}. In Table-I, II and III,  we tabulated the calculated values of the center and surface redshift  for the compact objects SAX J1808.4-3658, 4U 1538-52 and Her X-1 by taking the same values which we have used for graphical representation in Fig.\,\ref{f1}.

\section{Physical features and stability analysis of Compact Objects}

To gain some insight into the electrically charged fluids stellar model, we perform some analytical calculations and studied physical properties of the interior of the fluid sphere. The structure of charged spheres are analyzed by plotting several figures and studied equilibrium conditions under different forces. The obtained solution in this paper is used to study relativistic compact stellar objects within specified observational constraint.

\subsection{Tolman-Oppenheimer-Volkoff (TOV) equations}
To address the question, how gravitational and other different of fluid forces counteract with increasing electrostatic repulsion when the pressure gradients tend to vanish towards the boundary and charged fluid becomes more diluted.This situation may be illustrated by considering the hydrostatic equilibrium under different forces. This can be easily achieved by adapting generalized Tolman-Oppenheimer-Volkoff (TOV) equation \cite{35,36} in the presence of charge, as prescribed by Ponce de Le´on \cite{Ponce} is
\begin{eqnarray}
 -\frac{M_G(\rho+p)}{r^2}e^{\frac{\lambda-\nu}{2}}-\frac{dp}{dr}+
 \sigma \frac{q}{r^2}e^{\frac{\lambda}{2}} =0,\label{27} 
 \end{eqnarray}
where $M_G = M_G(r)$ is the effective gravitational mass inside a sphere of radius r
and q = q(r) is given by (\ref{14}). The expression for the effective gravitational mass is given by 
\begin{eqnarray}
M_G(r)=\frac{1}{2}r^2 \nu^{\prime}e^{(\nu - \lambda)/2}.\label{28}
\end{eqnarray}
Now, plugging the value of $M_G(r)$ in Eq. (\ref{27}), we get
\begin{eqnarray}
-\frac{\nu'}{2}(\rho+p)-\frac{dp}{dr}+\sigma \frac{q}{r^2}e^{\frac{\lambda}{2}} =0,  \label{29}
\end{eqnarray}

\begin{figure}[h!]
\centering
\includegraphics[width=7cm]{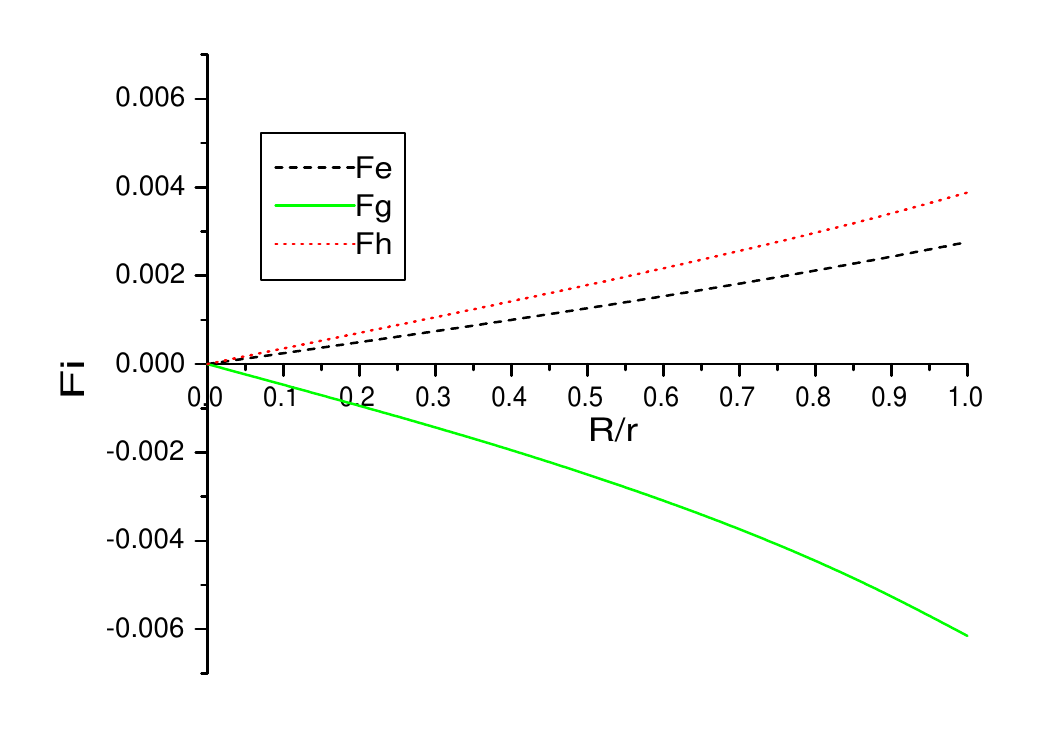} \includegraphics[width=7cm]{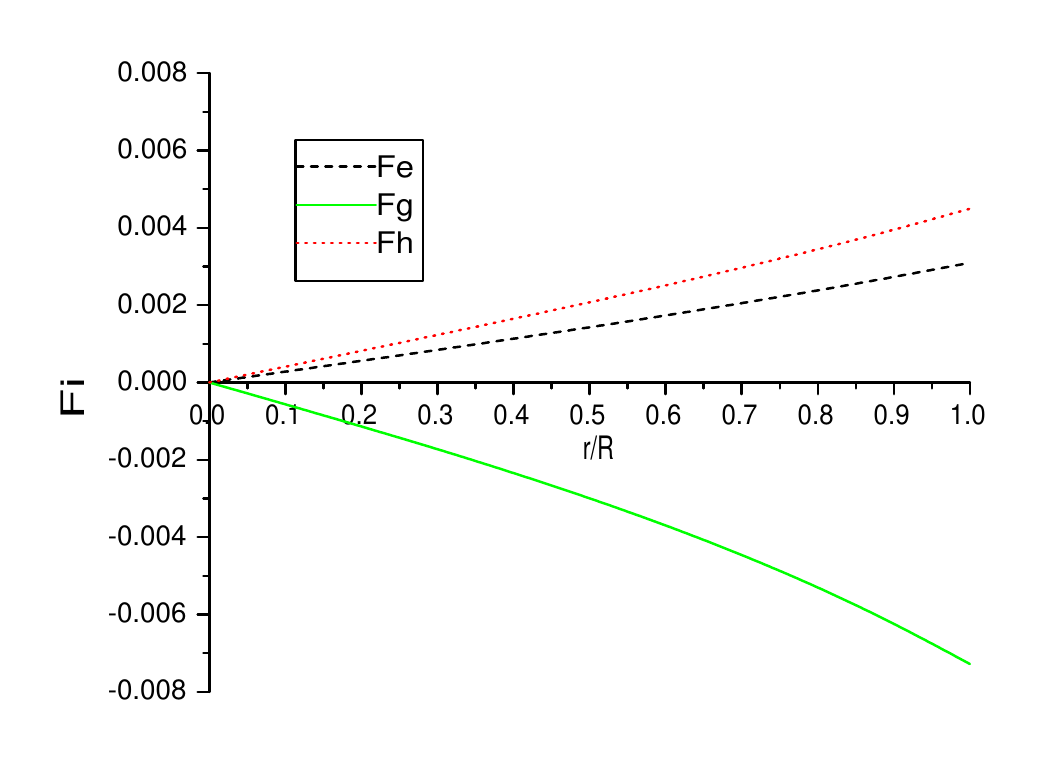}\\
\includegraphics[width=7cm]{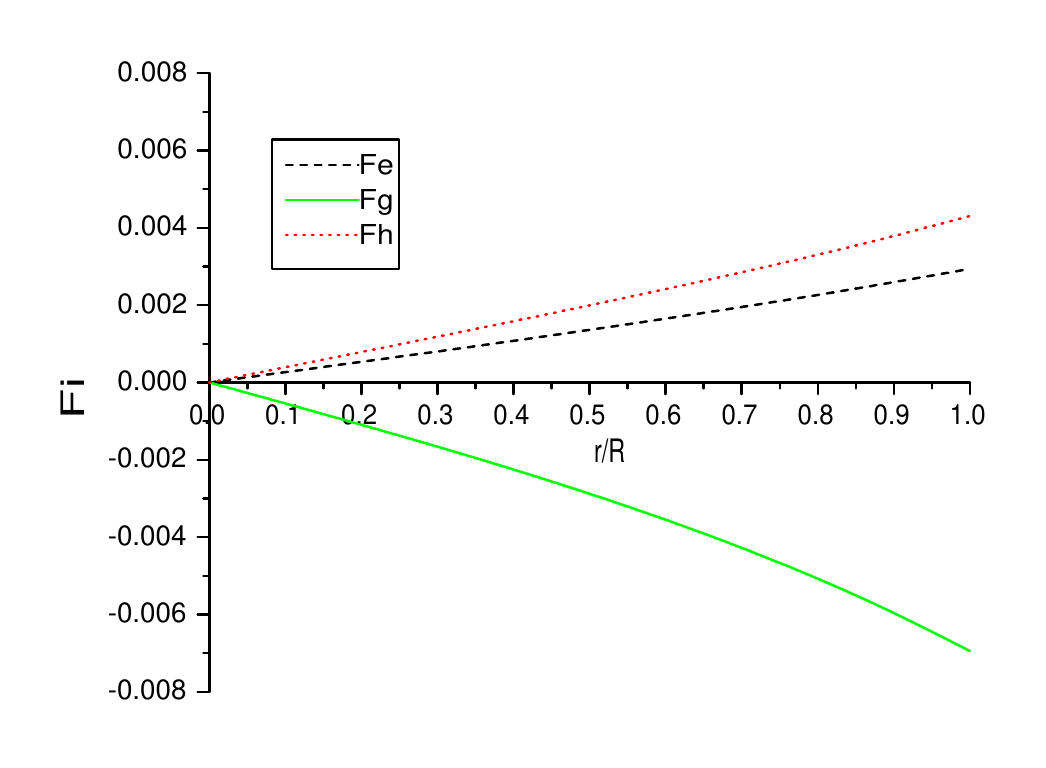}
\caption{\emph{We have plotted different forces, namely,  gravitational $(F_g)$, hydrostatic $(F_h)$ and electric forces $(F_e)$, respectively, to describe the equilibrium condition for charged fluid sphere. The result shows that gravitational force is dominated by hydrostatic and electric forces to maintain the equilibrium condition. We follow the same procedure for finding solutions as given in Fig.\,\ref{f1}.}}
\label{f2}
\end{figure}

Eq. (\ref{29}), gives the information about the stellar equilibrium configuration for charged relativistic fluid, subject to the gravitational $(F_g)$, hydrostatic $(F_h)$ and electric forces $(F_e)$, respectively, which are defined as: 
\begin{eqnarray}
F_g(r) = -\frac{\nu'}{2}(\rho+p)=\dfrac{Z'}{8\pi Z}(\rho+p),\label{30}
\end{eqnarray}
\begin{eqnarray}
F_h(r) = -\frac{dp}{dr}=-\dfrac{1}{8\pi}\left[{\dfrac{N}{(N1.N4)^{2}}-\dfrac{2\sqrt{Cx}}{K(1+Y^{2})^{2}}+M3.M4+M5.M6}\right],  \label{31}
\end{eqnarray}
\begin{eqnarray}
F_e(r) = \sigma \frac{q}{r^2}e^{\frac{\lambda}{2}}= \frac{1}{8\,\pi\,r^4}\,\frac{dq^2}{dr}=\dfrac{1}{16\pi}\left[\dfrac{\sqrt{Cx}[\theta^{2}(4-3K+x)-3b(1-K)]}{4(1-K)^{2}(1+Y^{2})^{7/4}}+\dfrac{\theta^{2}\sqrt{Cx}}{(1-K)(1+Y^{2})^{3/4}}\right].\label{32}
\end{eqnarray}
In Fig.\,\ref{f2},  the behavior of these forces for the onset of hydrostatic equilibrium are shown for the compact star candidates SAX J1808.4-3658, 4U 1538-52 and Her X-1. It is worth mentioning that, the whole system
is counterbalanced by the components of gravitational
force $(F_g)$, hydrostatic force$(F_h)$ and electric
force $(F_e)$ and the system attains a static equilibrium.

\subsection{Stability Analysis}
Now, we are interested in analysing, the speed of sound propagation $v^2_s $, which is given by the expression $v^2_s = dp/ d\rho$. In natural the velocity of sound  is less than the velocity of light. Here one can consider the speed of light is $c = 1$, so the sound speed is always less than unity. At this stage we  investigate the sound speed for charged fluid matter and for stable equilibrium configurations this should  belongs to the interval $0 < v^{2}=\frac{dp}{d\rho} < 1$,  as in ref. \cite{Herrera(2016)} for a subluminal sound speed. We  take Eq. (\ref{20}) and  (\ref{21}) for obtaining an explicit solution, which is
\begin{eqnarray}
\dfrac{dp}{c^{2}d\rho}=\dfrac{\dfrac{N}{(N1.N4)^{2}}-\dfrac{2\sqrt{Cx}}{K(1+Y^{2})^{2}}+M3.M4+M5.M6}{\bigg(N7-N8-N9\bigg)},\label{33}
\end{eqnarray}

\begin{figure}[h!]
\centering
\includegraphics[width=7cm]{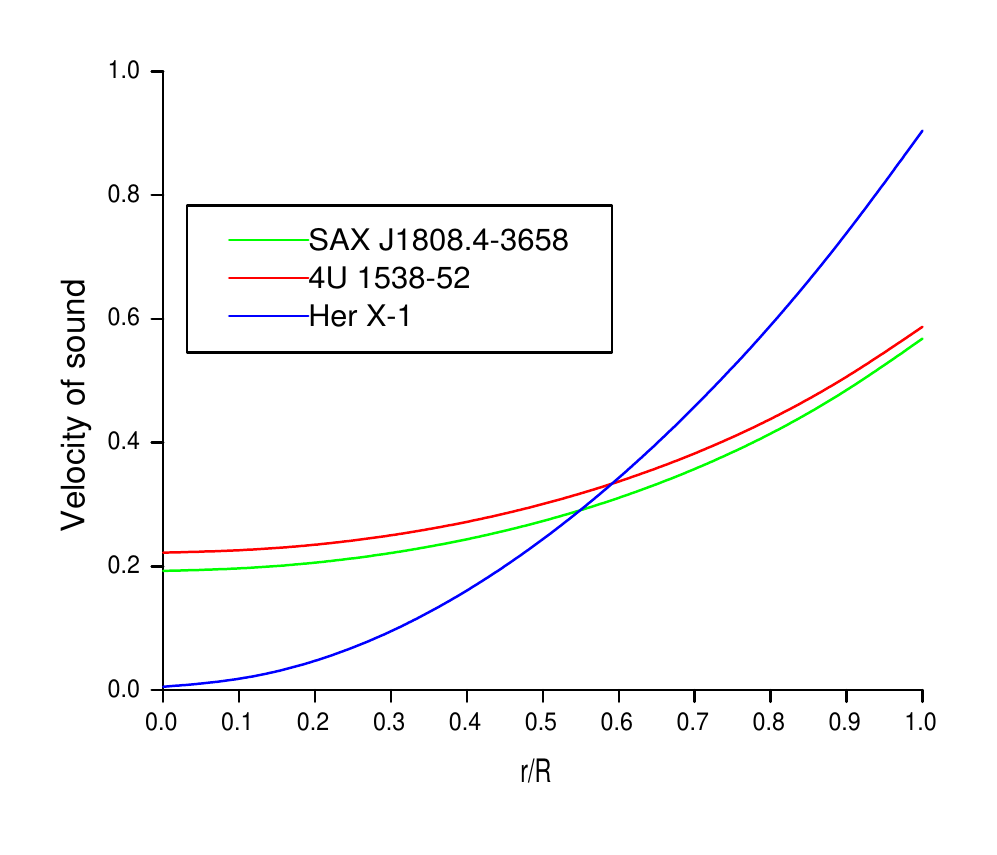} 
\caption{\emph{We have plotted the velocity of sound  against r/R for the compact star candidates SAX J1808.4-3658, 4U 1538-52 and Her X-1 inside the stellar interior by employing the same values of the constants as mentioned in Fig.\,\ref{f1}.}}
\label{f3}
\end{figure}
where we use for notational simplicity \\
$  N = (N1.N4).M1.(N1.N2+N3)+M7\bigg[(N1.N2)\bigg(M8.N2+N1(M8+M9)+M10.M11+M12.M13\bigg)-(N1.N2+N3)(M14.N4+N1.M15)\bigg],\\ \\ 
\text{with}~~~ M1=\dfrac{4\sqrt{Cx}}{K(1+Y^{2})^{2}},~~~~~~~M3=\dfrac{\sqrt{Cx}(1-x)}{K(1+x)^{3}},\\ \\ \\
M4=\bigg[K-1+\dfrac{2-3Y^{2}}{4(1+Y^{2})^{2}}+\dfrac{2\theta^{2}(1+Y^{2})}{(Y^{2}\theta^{2}+b)}\bigg],~~~~~~~M5=\dfrac{x}{2K(1+x)^{2}},\\ \\ \\
M6=\dfrac{5(K-1)\sqrt{Cx}}{2K(1+x)^{2}}-\dfrac{-2\theta^{2}}{(Y^{2}\theta^{2}+b)}\dfrac{2\sqrt{Cx}}{K(1-K)}-\dfrac{(1+Y^{2})}{K}\dfrac{4\sqrt{Cx}\theta^{4}}{(1-K)(Y^{2}\theta^{2}+b)^{2}},\\ \\ \\
M7=\dfrac{Y^{2}}{K(1+Y^{2})},~~~~~~M8=\dfrac{\sqrt{Cx}[\theta^{2}(4-3K+x)-3b(1-K)]}{4(1-K)^{2}(1+Y^{2})^{7/4}},~~~~~M9=\dfrac{\theta^{2}\sqrt{Cx}}{(1-K)(1+Y^{2})^{3/4}}, \\ \\ \\
M10=\dfrac{\sqrt{Cx}(K-2-x)}{4b(1-K)^{3/2}(K+x)^{3/2}(1+Y^{2})^{3/4}},~~~M11=1+\cos2\bigg(\arctan\bigg( \dfrac{Y}{\sqrt{b}}\theta\bigg)\bigg),\\ \\ \\
 M12=\dfrac{(1+Y^{2})^{1/4}}{2b(1-K)Y},~~~~~~M13=\dfrac{-4\sqrt{Cx} \theta}{b(1-K)+(K+x)\theta^{2}}\sin2\bigg(\arctan\bigg( \dfrac{Y}{\sqrt{b}}\theta\bigg)\bigg),\\ \\ \\
 M14=\dfrac{2\sqrt{Cx}}{(1-K)(1+Y^{2})^{3/4}}[1/4(Y^{2}\theta^{2}+b)+\theta^{2}(1+Y^{2})],\\ \\ \\
 M15=\dfrac{\sqrt{cx}}{2Yb(b(1-K)+(K+x)\theta^{2})} \left[1+\cos2\bigg(\arctan\bigg(\dfrac{Y}{\sqrt{b}}\theta\bigg)\bigg)\right],~~~N8=M3.M4,\\ \\ \\
N9=M5.M6,~~~~~N7=\dfrac{2\sqrt{Cx}(1-K)(5+x)}{K(1+x)^{3}}. $

In spite of these expression complexity, we use the graphical representation to represent it more conveniently. Using the expressions for all the terms in this formula, we have plotted Fig.\,\ref{f3}. From Fig.\,\ref{f3}, it is clear that the velocity of sound lies within the proposed interval and therefore our model maintains stability. Our investigation show that our proposed model for charged perfect fluid star satisfies both energy and stability conditions.

\subsection{Energy conditions}

To clarify the question of whether the present model satisfy all the energy conditions within the framework of general relativity or not. Clearly, such configurations depend on the relationship between matter density and pressure obeying certain restrictions. In connection with that, energy conditions are essential tools to understand many theorems of classical general relativity -such as the singularity theorems of stellar collapse.   Basic definitions are given for instance in \cite{Hawking,Visser}. Among them, we will be particularly interested in the (i) the Null energy condition (NEC), (ii) Weak energy condition (WEC) and (iii) Strong energy condition (SEC). Such conditions have the following inequalities
\begin{subequations}
\label{34}
\begin{align}
 \rho(r)+p \geq  0,\label{subeq1}\\
  \rho(r)+ \frac{q^2}{8 \pi r^4} \geq  0,\label{subeq2}\\
\rho+p+\frac{q^2}{4 \pi r^4} \geq  0, \label{subeq3}\\
\rho+3\,p +\frac{q^2}{4 \pi r^4}\geq  0.\label{subeq4}
\end{align}
\end{subequations}

\begin{figure}[h!]
\centering
\includegraphics[width=7cm]{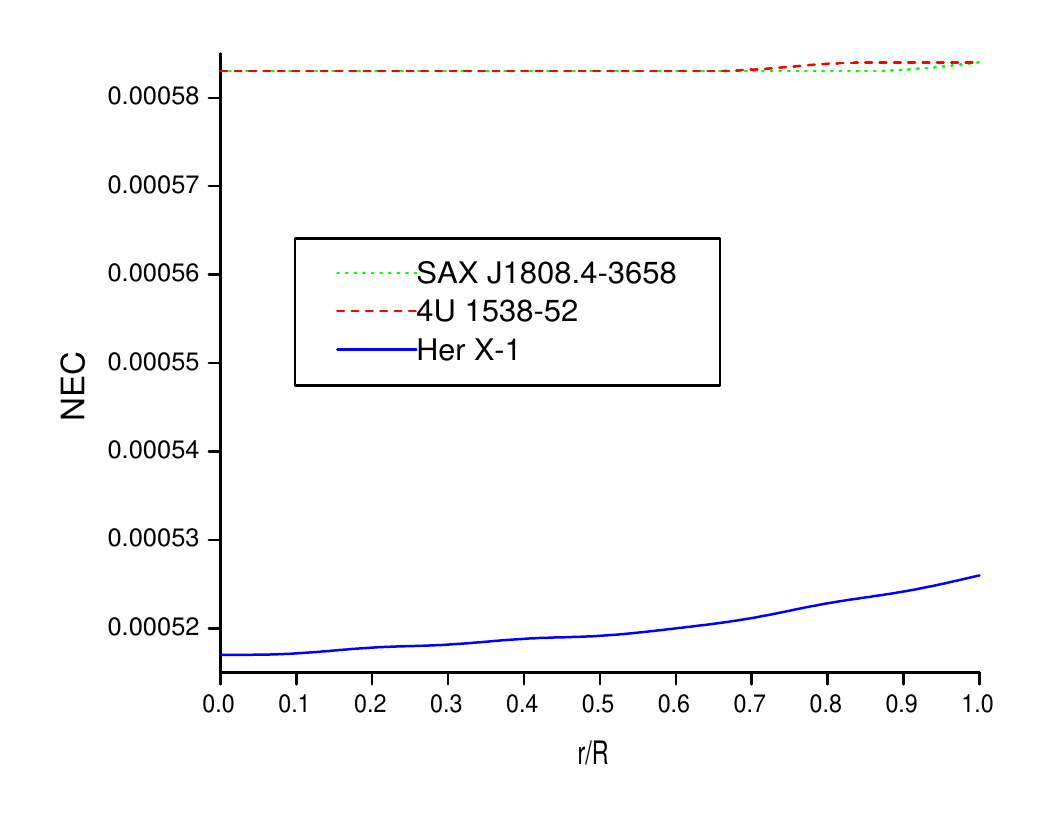} \includegraphics[width=7cm]{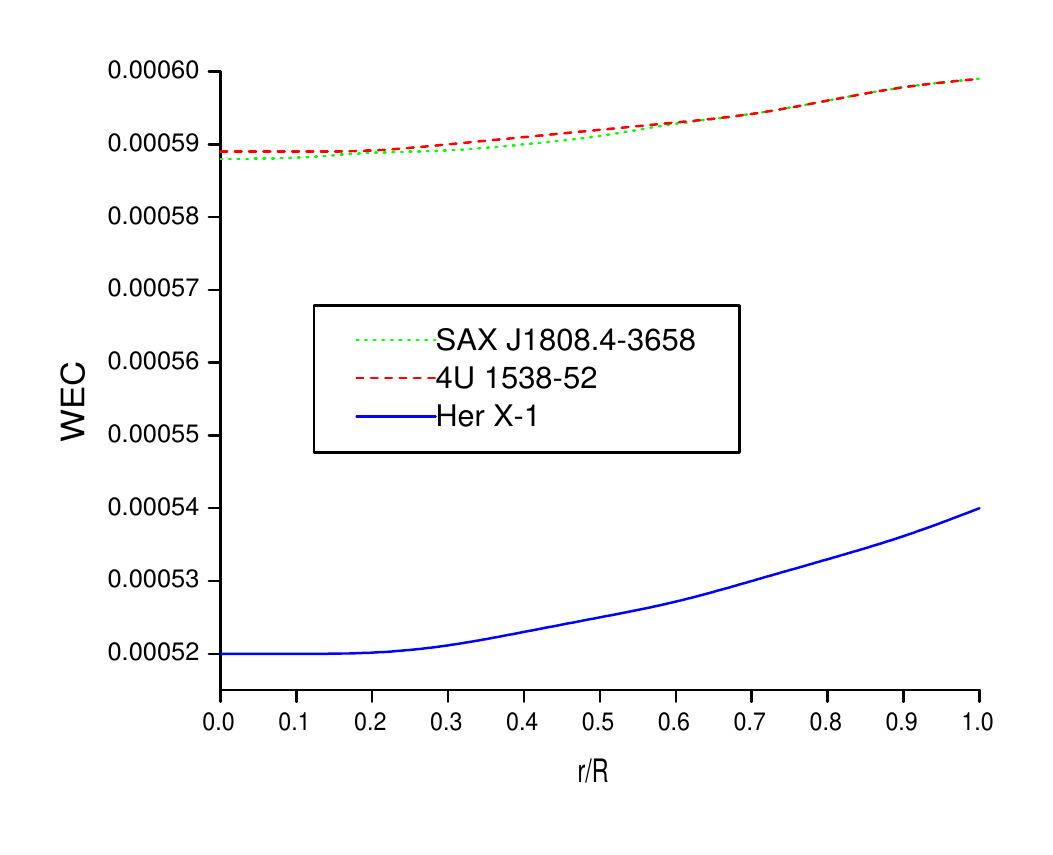}\\
\includegraphics[width=7cm]{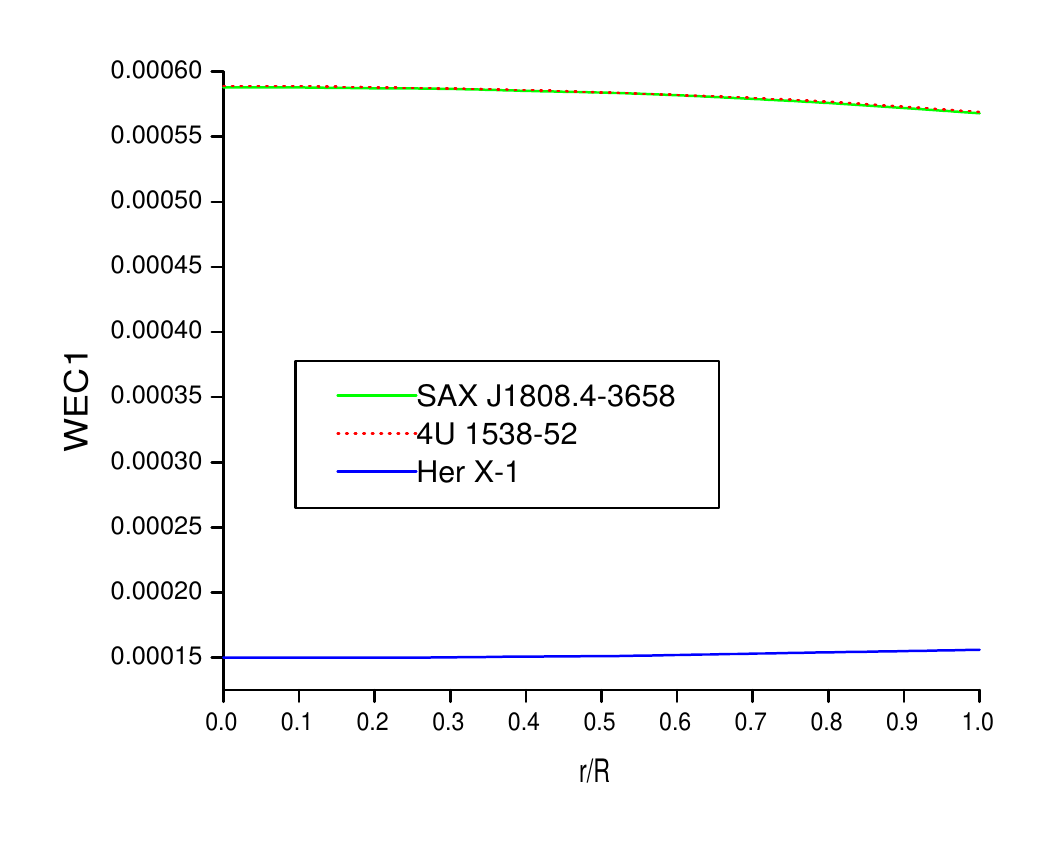}{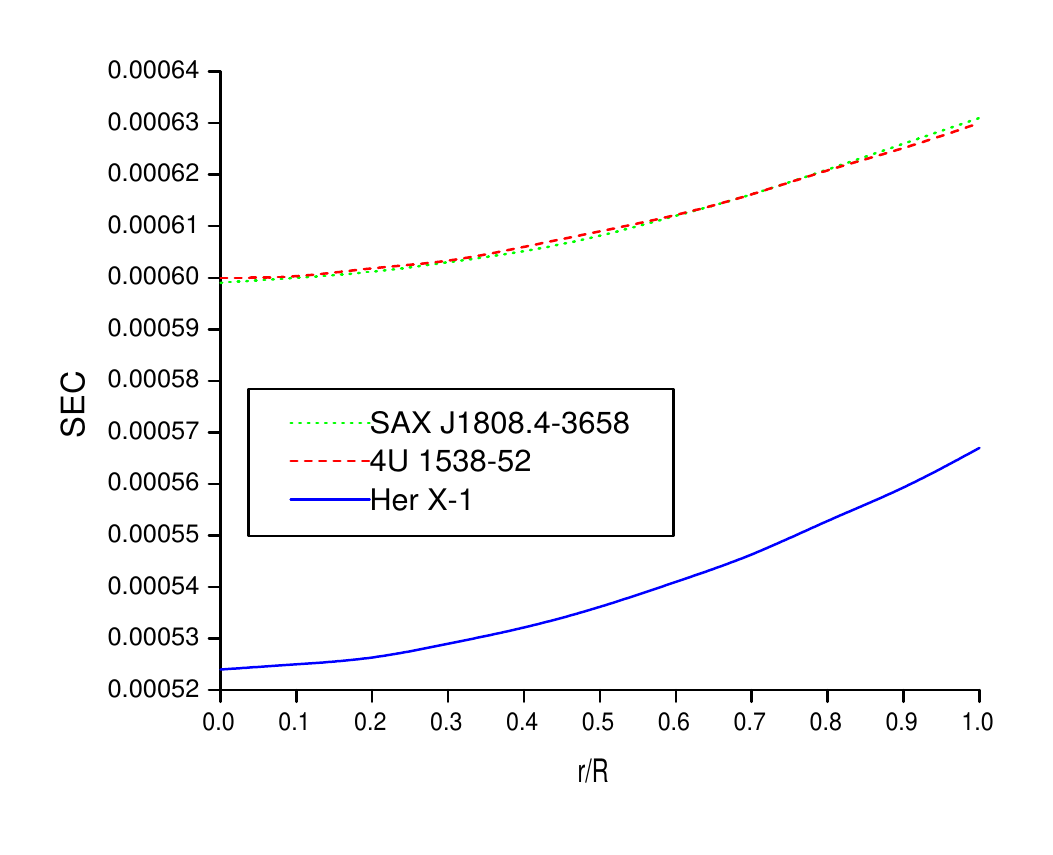} 
\includegraphics[width=7cm]{sec.PDF} 
\caption{\emph{Curves are plotted for the NEC, WEC, and SEC for the compact objects SAX J1808.4-3658, 4U 1538-52 and Her X-1. For the purposes of this calculation, we use the same values as given in Fig.\,\ref{f1}.}}
\label{f4}
\end{figure}
Inequalities (\ref{34}) hold automatically for the sources considered here. The weak energy condition imposes the requirement of a positive energy density as measured by a distant observer. Using this inequalities one can easily justify the nature of energy conditions for the specific stellar configuration as shown in Fig.\,\ref{f4}, that are satisfied for our proposed model.

\begin{center}\label{Table11-1}
\begin{tabular}{ |p{1.4cm}||p{2.1cm}|p{2.1cm}|p{2.1cm}|p{2.1cm}| p{2.1cm}|p{2.1cm}| } 
\hline \textbf{Table I.} & \multicolumn{6}{c|}{Values of the model parameters $K= 0.0032, C_{1} = - 0.00099,$}  \\
& \multicolumn{6}{c|} {$\theta^{2}$ = 1.006, b = 0.01, Zo = 0.02399, Za = 0.01595} \\ \hline
r/R & Pressure(P) & Density(D) &	Charge(q) & $dp/c^{2}d\rho$ &	P/D & $\gamma$ \\ \hline
0&	0.008805&	0.925155&		0&	0.192807&	0.009518&	20.450923\\ \hline
0.2&	0.008631&	0.924282&		0.019012&	0.205012&	0.009339&	22.158333\\ \hline
0.4&	0.008037&	0.921627&		0.152807&	0.242676&	0.008721&	28.069734\\ \hline
0.6&	0.006784&	0.917083&		0.519778&	0.309351&	0.007398&	42.126409\\ \hline
0.8&	0.004398&	0.910459&		1.245897&	0.412643&	0.00483&	85.846121\\ \hline
1&	0&	0.901461&	0.901461&		0.568269&	0&	Inf \\ \hline
\end{tabular}
\label{T1}
\end{center}

\newpage
\begin{center}\label{Table11-2}
\begin{tabular}{ |p{1.5cm}||p{2.1cm}|p{2.1cm}|p{2.1cm}|p{2.1cm}| p{1.9cm}|p{1.9cm}| }
\hline \textbf{Table II.}& \multicolumn{6}{c|}{Values of the model parameters $ K = 0.0033, C_{1} = -0.001, $}  \\
& \multicolumn{6}{c|} {$\theta^{2} = 4.83$, b = 0.05, Zo = 0.307064, Za = 0.197225} \\ \hline
r/R & Pressure (P) & Density (D) & 	Charge (q) & $dp/c^{2}d\rho$ &	P/D & $\gamma$ \\ \hline
0&	0.008896&	0.906091&		0&	0.222262&	0.009818&	22.861536\\ \hline
0.2&	0.008706&	0.905262&		0.018368&	0.234155&	0.009617&	24.580939\\ \hline
0.4&	0.00807&	0.902741&		0.147609&	0.270851&	0.00894&	30.568676\\ \hline
0.6&	0.006764&	0.898431&		0.501988&	0.335796&	0.007528&	44.940038\\ \hline
0.8&	0.004344&	0.892153&		1.202877&	0.436318&	0.00487&	90.038678\\ \hline
1&	0&	0.883639&	0.883639&		0.587405&	0&	Inf \\ \hline
\end{tabular}
\label{T2}
\end{center}

\begin{center}\label{Table11-3}
\begin{tabular}{ |p{1.2cm}||p{2.1cm}|p{2.1cm}|p{2.1cm}|p{2.1cm}| p{2.1cm}|p{2.1cm}| }
\hline \textbf{Table III.}& \multicolumn{6}{c|}{Values of the model parameters $ K = 0.034, C_{1} = -0.001, $}  \\ 
& \multicolumn{6}{c|} {$\theta^{2}$ = 0.835, b = 0.09, Zo = 0.284819, Za = 0.184272,} \\ \hline
r/R & Pressure (P) & Density (D) & 	Charge (q) & $dp/c^{2}d\rho$ &	P/D & $\gamma$ \\ \hline
0&	0.00397&	0.852353&		0&	0.005387&	0.004657&	1.161964\\ \hline
0.2&	0.003963&	0.852097&		0.017286&	0.044347&	0.004651&	9.578169\\ \hline
0.4&	0.003882&	0.851298&		0.139004&	0.158276&	0.00456&	34.868049\\ \hline
0.6&	0.003521&	0.849862&		0.473236&	0.340569&	0.004143&	82.548132\\ \hline
0.8&	0.002475&	0.847622&		1.135694&	0.586951&	0.00292&	201.613295\\  \hline
1&	0&	0.844317&	0.844317&		0.904533&	0&	Inf  \\ \hline
\end{tabular}
\label{T3}
\end{center}

\begin{center}\label{Table11-4}
\begin{tabular}{ |p{3.2cm}||p{1.7cm}|p{1.7cm}|p{2.5cm}|p{2.4cm}| p{2.4cm}|p{1.7cm}| }
\hline \textbf{Table IV.} & \multicolumn{5}{c|}{The numerical values of the masses-radius (M/R) and  }  \\
& \multicolumn{5}{c|} {Surface density $(\rho_{S})$ for the compact star candidates} \\ \hline

\text{Compact Stars} & ~R (Km.) & ~M($M_{\odot})$ & - C ($Km^{-2} $) & $\rho_{S}$ ($g/cm^{3}$) & M/R  \\ \hline
4U 1538-52  & 7.866   & 0.87 & 1.6162$ \times 10^{-5}$ & 7.66$ \times 10^{14}$ & 0.16314 \\ \hline

Her X-1  & 8.1   & 0.85 &  1.5089$ \times 10^{-5} $ & 6.918$ \times 10^{14}$ & 0.1547  \\ \hline

SAX J1808.4-3658 & 7.951   & 0.9 &  1.5659$ \times 10^{-5}$   & 7.6490$ \times 10^{14}$  & 0.16696  \\ \hline
\hline
\end{tabular}
\end{center}

\begin{figure}[h]
\centering
\includegraphics[width=7cm]{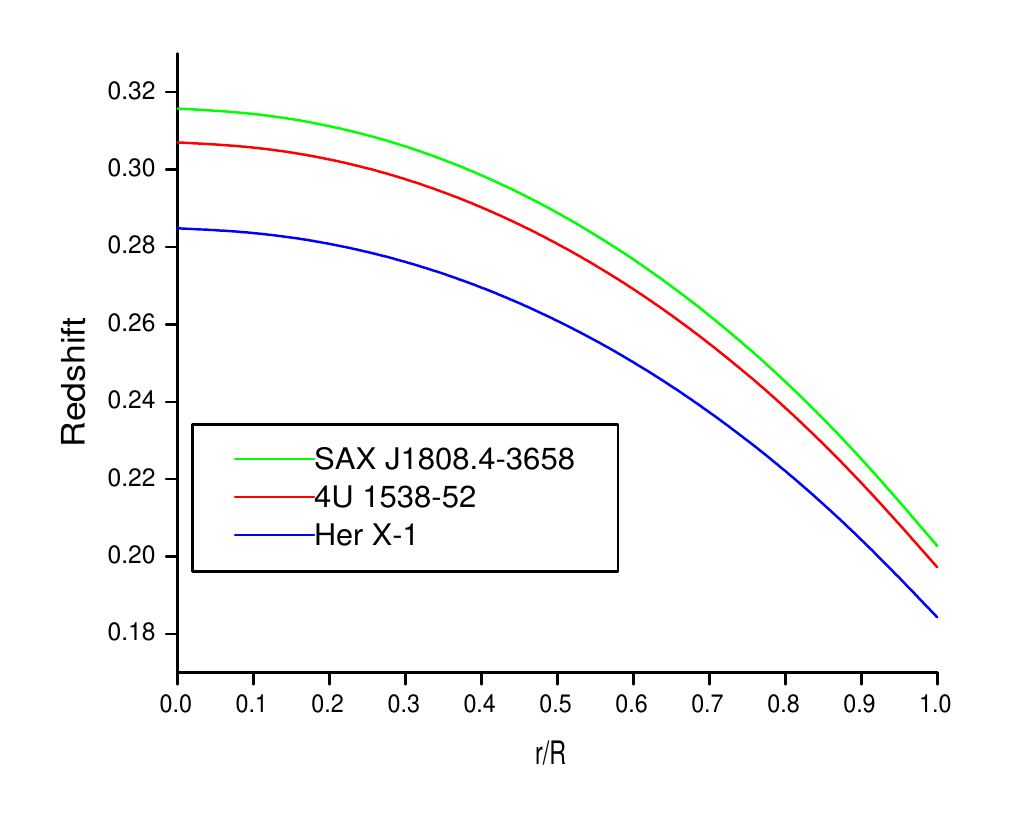} \includegraphics[width=7cm]{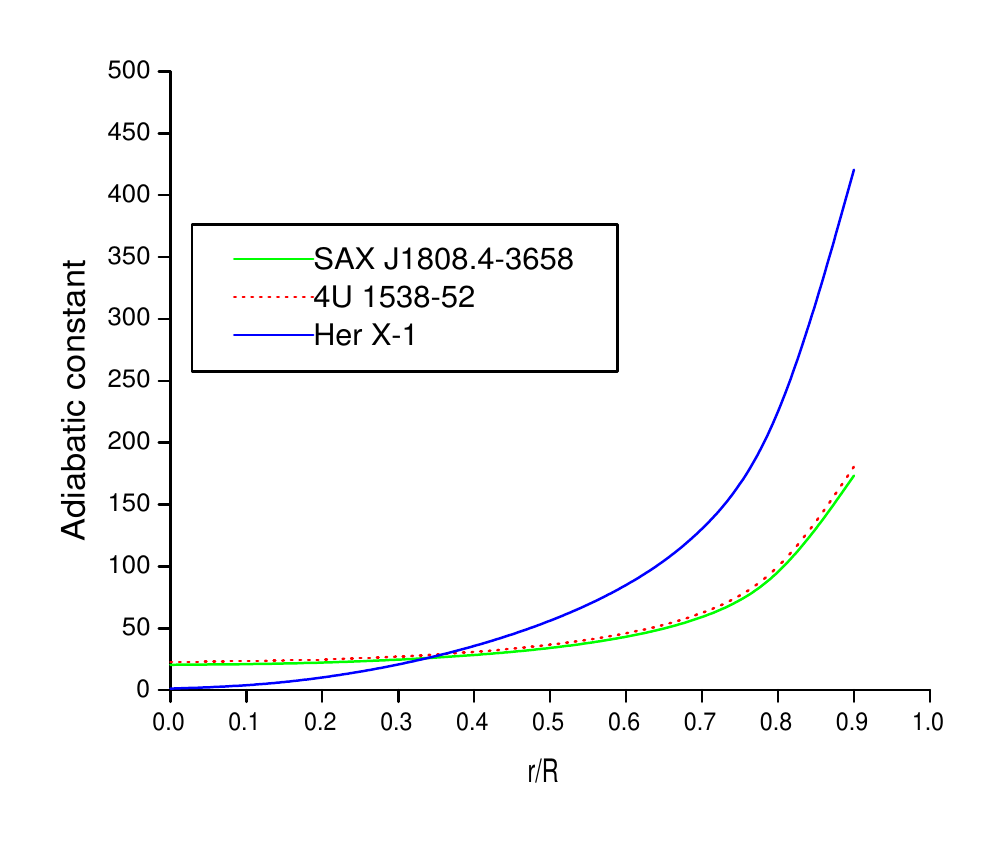}
\caption{\emph{Behaviour of redshift and adiabatic constant vs. fractional radius
r/R for SAX J1808.4-3658, 4U 1538-52 and Her X-1. For plotting we have employed data set values of physical parameters and constants which are the same as used in Fig.\,\ref{f1}}}
\label{f5}
\end{figure}

\subsection{Relativistic adiabatic index and Stability}
We have so far considered stars in both hydrostatic equilibrium and  speed of sound.  But an important question that remains to be answered is whether these centenarians are enough for stable analysis. As another application is the adiabatic index. Using this one can incorporate all the basic characteristics of an equation of state on the instability formulae.  In the case of an EOS of neutron star matter, the adiabatic index  $\gamma$, varies from 2 to 4 and for an anisotropic compact star will be stable if $\gamma > 4/3$ \cite{Heintzmann}. For an adiabatic perturbation, the adiabatic index  $\gamma$, can be expressed in the form \cite{Merafina}.

\begin{eqnarray}
\gamma =\left(\left(\dfrac{c^{2}\rho+p}{p}\right)\left(\dfrac{dp}{c^{2}d\rho}\right)\right) > 4/3.\label{35} 
\end{eqnarray}
Starting with Eqs. (\ref{20}), (\ref{21})  and (\ref{35}), one can easily justify the adiabatic index and the stable configuration
for the specific stellar configuration SAX J1808.4-3658, 4U 1538-52 and 
Her X-1. In order to clarify further we plot in Fig.\,\ref{f5}, the dependence of the averaged adiabatic index, which clearly indicates that
the configurations developed in this paper is stable. Finally, it should be noted that the adiabatic index which we define in Eq. (\ref{35}), is a local characteristic of a specific EOS and depends on interior fluid density. For further reference we refer our reader to follow \cite{Moustakidis}.

\section{Conclusion}
The main point we wish to make is that  a general solution of the Vaidya-Tikekar model for a spherically symmetric superdense star has been carried out, and it is found a physically valid solution. The spheres contain a charged perfect fluid matter and the exterior spacetime is represented by the Reissner-Nordstr$\ddot{\text{o}}$m metric. We also intend to study more closely the physical features of compact stars, generating an exact solution, and compated with number of specific astronomical objects. We summarize the obtained results for the interior of the stellar configuration as follows:\\
\\
(i)\textit{ Regularity conditions}: The energy density  and pressure is positive inside the star. The central density $\rho_{0}= \left(3C(K-1)\right)/K > 0$ and central pressure $ p_{r}(r=0)>0$, which shows that the density and pressure are positive inside the star.\\
(ii) \textit{Generalized TOV equation}: We consider the generalized TOV equation for desribing the equilibrium condition subject to gravitational (F$_g$), hydrostatic (F$_h$) and electric forces (F$_e$), respectively and we observe from Fig.\,\ref{f2}, that gravitational force is balanced by the joint action of hydrostatic and electric forces to attain the required stability of the model. However, the effect of electric force is less than the hydrostatic force.\\
(iii) \textit{Causality conditions}: In addition, we have shown that inside the charged fluid sphere the speed of sound is less than the speed of light i.e. $0< (dp/c^{2}d\rho)<1 $ for different values of $0< K <1$ (see Table I-III for more detail results). However, we can see from Fig.\,\ref{f3} that the sound speed is increasing monotonically towards the surface of the star.\\
(iv) \textit{Energy conditions}: The configurations of the objects satisfy the null energy condition (NEC), weak energy condition (WEC) and strong energy condition (SEC), simultaneously inside the star as displayed in Fig.\,\ref{f4}. \\
(v) \textit{Surface Redshift}: We have also studied another impotent features is the surface redshift of stellar structure. A general feature of the behavior is that it attend maximum value at the center and monotonically decreasing towards the boundary, which we can see from figure Fig.\,\ref{f5} (left panel). The numerical values corresponding to center and surface redshift for the SAX J1808.4-3658, 4U 1538-52 and  Her X-1 are enlisted in Table I-III.\\
(vi) \textit{Stability conditions}: We have also considered the adiabatic index to gain some useful information at an extreme conditions and for stable configuration of a compact object. In this case it is possible to approximation  the adiabatic index $\gamma >$ 4/3. We demonstrate that in Tables I-III and graphical form as displayed in Fig.\,\ref{f5} (right panel).\\
(vii) \textit{ Mass-radius ratio}: As an another application of the obtained upper bound of mass-radius relation together with the  Buchdahl limit  for perfect fluids  satisfying the inequality $R\leq$ (9/8) $R_S$ =  (9/4) G M/$c^2$ \cite{Buchdahl}. For a compact configuration with electric charge to the system, the Buchdahl limit is certainly increase. It has been found that Buchdahl limits for charged stars in \cite{Yunqiang,Dobson,Giuliani}. In our case the obtained   mass-to-radius ratio fall within the bounds as proposed for charged spheres (see Table -4 for more).

In conclusion, we will  investigate other forms of metric potentials that could exhibit more general behavior and to make observationally distinguish between these compact objects. \\
\\
\textbf{Acknowledgments}:
AB wishes to thank the University of KwaZulu-Natal (ACRU) for financial support.


\begin{thebibliography}{12}
\bibitem{Vaidya} P. C. Vaidya and R. Tikekar: {\it  J. Astrophys. Astro.}, {\bf 3}, 325 (1982).
\bibitem{Mak} M. K. Mak, P. N. Dobson, Jr. and T. Harko: {\it Mod.Phys.Lett. A}, {\bf15}, 2153-2158 (2000). 
\bibitem{Burikham} P. Burikham \textit{et al:} {\it Eur.Phys.J. C}, {\bf75}, 442 (2015). 
\bibitem{Boehmer1} C. G. Boehmer and T. Harko: {\it  Gen.Rel.Grav.}, {\bf39},  757-775 (2007). 
\bibitem{Ray} Subharthi Ray \textit{et al:} {\it Braz.J.Phys.}, {\bf 34}, 310-314 (2004).
\bibitem{Negreiros} R. P. Negreiros \textit{et al:} {\it Phys.Rev. D}, {\bf 80}, 083006 (2009). 
\bibitem{Varela} Victor Varela \textit{et al:} {\it Phys.Rev. D}, {\bf 82}, 044052 (2010).
\bibitem{Maharaj1} S.D. Maharaj and P.Mafa Takisa: {\it Gen.Rel.Grav.}, {\bf 44}, 1419-1432 (2012) .
\bibitem{Hessels} J. Hessels \textit{et al:} {\it in Binary Radio Pulsars, PASP Conference Series}, {\bf 328}, 395 (2005).
\bibitem{Bower} R. L. Bower and E. P.T Liang: {\it  Astrophys. J.}, {\bf 188}, 657 (1974).
\bibitem{Kalam1} Mehedi Kalam \textit{et al:} {\it Astrophys.Space Sci.}, {\bf 349}, 865-871 (2014).
\bibitem{Kalam2} Mehedi Kalam \textit{et al:} {\it Eur.Phys.J. C}, {\bf 73},  2409 (2013).
\bibitem{Rahaman1}  Farook Rahaman \textit{et al:} {\it Eur.Phys.J. C}, {\bf 72}, 2071 (2012). 
\bibitem{Bhar1} P. Bhar, M. H. Murad and N. Pant: {\it Astrophys. Space Sci.}, {\bf 359}, 13 (2015).
\bibitem{Bhar2} P. Bhar \textit{et al:} {\it Eur.Phys.J. A}, {\bf 52}, 312 (2016).

\bibitem{Banerjee1} S.K. Maurya, Ayan Banerjee and Phongpichit Channuie: e-Print: arXiv:1711.03412 [gr-qc].
\bibitem{Banerjee2}  S.K. Maurya \textit{et al:} {\it Annals Phys.}, {\bf 385}, 532-545 (2017).
\bibitem{Ratanpal}  B. S. Ratanpal, V. O. Thomas and D. M. Pandya: {\it Astrophys. Space Sci.}, {\bf 361}, 65 (2016).

\bibitem{Thirukkanesh} S. Thirukkanesh and F.C. Ragel: {\it Chin.Phys. C}, {\bf 41}, 015102 (2017).

\bibitem{Jose1}  Jos$\acute{e}$ D.V. Arbañil and Manuel Malheiro: {\it AIP Conf.Proc.}, {\bf 1693}, 030007 (2015). 
\bibitem{Jose2} Jos$\acute{e}$ D. V. Arbañil and M. Malheiro : {\it Phys.Rev. D}, {\bf 92},  084009 (2015).
\bibitem{Maurya3}  S.K. Maurya, Y.K. Gupta, S. Ray, S. R. Chowdhury : {\it Eur.Phys.J. C}, {\bf 75},  389 (2015).
\bibitem{Maharaj2}  S.D. Maharaj and P.Mafa Takisa: {\it  Gen.Rel.Grav.}, {\bf 44},  1419-1432 (2012).
\bibitem{Barreto1}  W. Barreto and L. Rosales: {\it Gen.Rel.Grav.}, {\bf 43 } 2833-2846 (2011).
\bibitem{Barreto2} W. Barreto \textit{et al}: {\it  Gen.Rel.Grav.}, {\bf 39},  537-538 (2007).
\bibitem{10} L. K. Patel and Kopper: {\it Aust. J. Phy}, {\bf 40}, 441 (1987).
\bibitem{11} R. Sharma, S. Mukherjee and S. D. Maharaj: {\it Gen. Rel. Grav.}, {\bf 33}, 999 (2001).
\bibitem{12} Y. K. Gupta and N. Kumar: {\it Gen. Rel. Grav.}, {\bf 37}, 575 (2005).
\bibitem{13} K. Komathiraj and S. D. Maharaj: {\it Int. J. Mod. Phys. D.}, {\bf 16}, 1803 (2007).
\bibitem{Patel} L. K. Patel and S. S. Koppar: {\it Aust. J. Phys}, {\bf 40},  441 (1987).
\bibitem{Bijalwan} N. Bijalwan and Y. K. Gupta: {\it Astrophys.Space Sci.} {\bf 334}, 293-229 (2011).
\bibitem{Bijalwan1} N. Bijalwan and Y. K. Gupta: {\it Astrophys. Space
Sci.}, {\bf 337}, 455–462 (2012). 
\bibitem{2013} J. Kumar and Y. K. Gupta: {\it Astrophys.Space Sci.}, {\bf 345}, 331–337 (2013).


\bibitem{2014} J. Kumar and Y. K. Gupta: {\it  Astrophys Space Sci}, {\bf 351}, 243–250 (2014). 
\bibitem{Gangopadhyay} T. Gangopadhyay \emph{et al.}: {\it Mon. Not. R. Astron. Soc.}, {\bf 431}, 3216 (2013).
\bibitem{Misner} C. W. Misner and D. H. Sharp: {\it Phys. Rev.}, {\bf 136}, B571 (1964).
\bibitem{Tikekar} R. Tikekar: {\it Journal of Mathematical Physics}, {\bf 31},  2454–2458, (1990).
\bibitem{Leach} S. D. Maharaj and P. G. L. Leach: {\it Journal of Mathematical Physics}, {\bf 37}, 430–437, (1996).
\bibitem{Florides} P. S. Florides: {\it J. Phys. A, Math. Gen.}, {\bf 17}, 1419 (1983).
\bibitem{Synge} J. Synge: {\it the general theory Series in physics} (North-Holland Pub. Co.), 1960 Relativity.
\bibitem{Buchdahl} H. A. Buchdahl: {\it Phys. Rev. D},  {\bf 116}, 1027, (1959).
\bibitem{Straumann} N. Straumann: {\it General Relativity and Relativistic Astrophysics} (Springer, Berlin, 1984) 43.
\bibitem{Karmakar}  S. Karmakar, S. Mukherjee, R. Sharma and S. D. Maharaj
: {\it Pramana }, {\bf 68}, 881 (2007).

\bibitem{Barraco}  D. E. Barraco, V. H. Hamity and R. J. Gleiser: {\it  Phys. Rev. D}, {\bf 67}, 064003 (2003).
\bibitem{Boehmer} C. G. Boehmer, T. Harko : {\it Class.Quant.Grav.}, {\bf 23},  6479-6491 (2006).
\bibitem{Ivanov} B. V. Ivanov: {\it  Phys. Rev. D}, {\bf 65}, 104001 (2002).
\bibitem{35} R. C. Tolman: {\it Phys. Rev.}, {\bf 55}, 364 (1939).
\bibitem{36} J. R. Oppenheimer and G.M. Volkoff: {\it Phys. Rev.}, {\bf 55}, 374
(1939).
\bibitem{Ponce} J. Ponce de Leon: {\it  Gen. Relativ. Gravit.}, {\bf 25}, 1123 (1993).
\bibitem{Herrera(2016)} L. Herrera: {\it Phys. Lett. A}, {\bf 165}, 206 (1992).
\bibitem{Hawking} S. W. Hawking and G. F. R. Ellis: {\it The large scale structure of space-time}, (Cambridge, England, 1973).
\bibitem{Visser} M. Visser: {\it Lorentzian wormholes—from Einstein to Hawking}, (AIP Press, New York, 1995).
\bibitem{Heintzmann} H. Heintzmann and W. Hillebrandt: {\it Astron. Astrophys. }, {\bf 38}, 51 (1975).
\bibitem{Merafina} M. Merafina and R. Ruffini: {\it Astron. Astrophys.}, {\bf 221}, 4 (1989).
\bibitem{Moustakidis} Ch.C. Moustakidis: {\it  Gen.Rel.Grav.}, {\bf 49} , 68 (2017).
\bibitem{Yunqiang} Y. Yunqiang and L. Siming, : {\it  Commun. Theor. Phys.}, {\bf 33}, 571 (2000).
\bibitem{Dobson} M. K. Mak, P. N. Dobson, and T. Harko : {\it Europhys. Lett.}, {\bf 55}, 310 (2001).
\bibitem{Giuliani} A. Giuliani and T. Rothman: {\it Gen. Relativ. Gravit.} , {\bf 40}, 1427 (2008).


\bibitem{Takisa} P. Mafa Takisa, S. D. Maharaj: {\it  Gen. Relativ. Gravit.}, {\bf 45}, {\bf 1951} (2013).
\bibitem{S1} R. Sharma, S. Mukherjee, and S. D. Maharaj: {\it  Gen. Relativ. Gravit.}, {\bf33}, 999, (2001).

\bibitem{Marahaj1} S. Thirukkanesh and S. D. Marahaj: {\it  Class. Quantum Grav.}, {\bf 25}, 235001 (2008).






\end{thebibliography}
\end{document}